\documentclass[twocolumn,prl,showpacs,amsmath,amstex,amssymb,mathfonts,superscriptaddress]{revtex4-1}
\pdfoutput=1
\usepackage{amsthm,color,amsfonts,graphicx,verbatim}
\usepackage{amsmath}
\usepackage{amssymb}
\usepackage{amsthm}
\usepackage{amsfonts}
\usepackage{listings}
\usepackage{enumerate}
\usepackage{latexsym}
\usepackage{psfrag}
\usepackage{bm}
\usepackage{graphicx}
\newcommand{\be}{\begin{equation}}
\newcommand{\ee}{\end{equation}}
\newcommand{\bea}{\begin{eqnarray}}
\newcommand{\eea}{\end{eqnarray}}

\renewcommand{\phi}{\varphi}
\renewcommand{\epsilon}{\varepsilon}
\renewcommand{\vec}[1]{{\bf #1}}

\begin{document}
\title{Spectral statistics across the many-body localization transition}
\author{Maksym Serbyn} 
\affiliation{Department of Physics, University of California, Berkeley, California 94720, USA}
\author{Joel E. Moore}
\affiliation{Department of Physics, University of California, Berkeley, California 94720, USA}
\affiliation{Materials Science Division, Lawrence Berkeley National Laboratory, Berkeley, California 94720, USA}
\date{\today}
\begin{abstract}
The  many-body localization transition (MBLT)  between ergodic and many-body localized phase in disordered interacting systems is a subject of much recent interest.  Statistics of eigenenergies is known to be a powerful probe of crossovers between ergodic and integrable systems in simpler examples of quantum chaos.  We consider the evolution of the spectral statistics across the MBLT, starting with mapping to a Brownian motion  process that analytically relates the spectral properties to the statistics of matrix elements. We demonstrate that the flow from Wigner-Dyson to Poisson statistics is a two-stage process. First, fractal enhancement of matrix elements upon approaching the MBLT from the metallic side produces an effective power-law interaction between energy levels, and leads to a plasma model for level statistics.  At the second stage, the gas of eigenvalues has local interaction and level statistics belongs to a semi-Poisson universality class.  We verify our findings numerically on the XXZ spin chain. We provide a microscopic understanding of the level statistics across the MBLT and discuss implications for the transition that are strong constraints on possible theories.
\end{abstract}
\maketitle
{\it Introduction.} Quantum and statistical mechanics represent two seemingly rather different approaches to the description of complex physical systems. Yet these two viewpoints agree for a wide class of isolated quantum systems, which are said to thermalize~\cite{SrednickiETH,RigolNature}. Determining the circumstances under which an isolated quantum many-body system becomes its own thermal bath and thermalizes itself, just as Baron Munchausen could pull himself out of a mire by his own hair, perhaps using some kind of fluctuation, is an open question. 

Phenomena similar to the emergence of thermalization also occur in few-body quantum systems, which frequently show the emergence of so-called quantum chaos~\cite{HaakeBook}. There, upon changing parameters/number of degrees of freedom, the classical system can go from regular to chaotic behavior. On a quantum level this results in changes of level statistics, which has proven to be a powerful probe of the system properties in the context of quantum chaos. In particular, there exist two standard universal limits: Poisson statistics (PS) and Wigner-Dyson level statistics (WDS)~\cite{MehtaBook}. For few-body systems, PS applies to systems which are classically integrable and do not have any level repulsion. WDS stems from random-matrix theory and holds for generic chaotic systems, where energy levels repel each other (i.e., the energy difference between neighboring levels is statistically unlikely to be small compared to the mean level spacing). 

Integrable (non-chaotic) behavior is abundant in the context of few-body physics. On the other hand, in the many-body world the only non-thermalizing \emph{phase} (in the sense of stability to small perturbations) is represented by many-body localized (MBL) systems~\cite{Basko06,Mirlin05}. Recent progress established that thermalization fails in the MBL phase due to the existence of extensively many conserved quantities~\cite{Huse13,Serbyn13-1,Imbrie14,ScardicchioLIOM}. On the other hand, it is known that one can tune the system through a phase transition into a thermalizing ergodic phase~\cite{PalHuse,Reichman14,Alet14,AltmanRG14,Potter15,Goold15,Luca13,Demler14,Santos15,Serbyn15}. Below we aim to understand the evolution of the level statistics across the MBL-to-ergodic transition, gaining insights into the breakdown of thermalization. 

Crossover between PS and WD statistics has been studied extensively in a single-particle physics context:  for quantum kicked rotor~\cite{Izrailev89}, integrability breaking perturbations~\cite{Haake91,Caurier}, and single-particle Anderson localization transition (ALT)~\cite{Shklovskii93,MirlinRMP,MirlinRPP}.  In the many-body problems, PS to WD crossover is also known to occur upon breaking of (quantum) integrability~\cite{Modak}. In most of the examples, the PS and WDS are the only two stable points. The only known exception is the ALT, where \emph{universal statistics} different from PS and WDS emerges at the mobility edge~\cite{Shklovskii93}.

The spectral statistics in the case of MBL transition was demonstrated to evolve from WDS to PS as one localizes the system~\cite{OganesyanHuse,PalHuse,Berkovits02,Modak14}, however not much is known about the intermediate statistics.  The common probe used to characterize level statistics across MBLT is an average ratio of the consecutive energy spacings~\cite{PalHuse,Alet14,Demler14,Reichman14}. However, this is a single parameter and it does not provide much insight into the intermediate form of the level statistic, nor into physical details of its crossover. 

In this paper we study how the spectral statistics changes across the MBL-delocalization transition. In order to build a microscopic understanding of the level statistics we generalize Dyson's Brownian motion model~\cite{Dyson62}, previously applied to the ALT~\cite{ChalkerRW}, to the many-body case. From the mapping to Brownian motion, we obtain non-trivial relations between fractality~\cite{Luca13,Demler14,Santos15,Serbyn15}, spectral statistics, and properties of matrix elements across the MBLT~\cite{Rahul14,Serbyn15}.  While many features can be simultaneously explained in this analysis, one surprise is that there appear to be two different regimes of intermediate spectral statistics: in one, the effective interaction between energy levels in the plasma model has a variable power-law, while in the other, the effective interaction is short-ranged but over a variable number of levels.

\begin{figure}[b]
\begin{center}
\includegraphics[width=0.95\columnwidth]{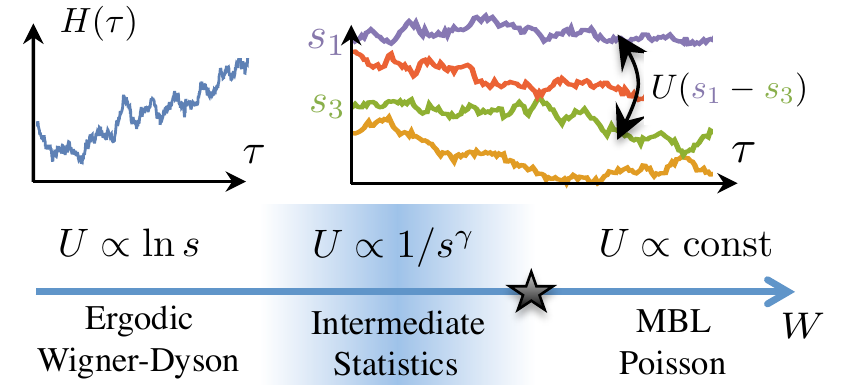}\\
\caption{ \label{Fig:cartoon} (top) Random walk in a space of Hamiltonians induces a stochastic process on the eigenenergies. The interaction between eigenlevels is set by a potential energy $U(s_i-s_j)$. (bottom) Evolution of the interaction between levels $U(s)$  across the MBL transition determines the level statistics.  }
\end{center}
\end{figure}

Within the picture of Brownian motion~\cite{Dyson62,ChalkerRW}, the level statistics is controlled by the effective interaction between energy levels, see Fig.~\ref{Fig:cartoon}. In particular, deep in the metal phase, the WD statistics emerges from the partition function of a one-dimensional Coulomb gas, where particles interact with a logarithmic potential $U(s) = \log |s|$.  At a first stage, upon approaching the MBL transition, the effective interaction starts to decay as a power-law: $U(s_i-s_j) =  |s_i-s_j|^{-\gamma}$ when $|s_1-s_2|\geq N_\text{erg}$. The power-law interaction changes tails of the level statistics, so it can be approximately described by the plasma model, and is intermediate between PS and WDS case.  
At the second stage, when exponent $\gamma$ becomes bigger than one,  the interaction becomes effectively short-ranged, and level spacing distribution tends to the semi-Poisson distribution~\cite{Bogomolny99}.  In this regime it is the range of the interaction which changes with disorder/system size. As soon as the range of interactions reaches zero, we arrive at Poisson statistics. 

Before discussing implications of the above picture of the level statistics, we justify the proposed cartoon using both analytic and numeric arguments. In particular, we argue that the parameter $\gamma$ introduced above can be extracted from the properties of the many-body matrix elements which decay as a power-law with energy separation between eigenstates, where $\gamma\leq 1$ is the same power which controls level statistics. The power-law behavior of matrix elements can be viewed as a generalization of the Chalker-Daniell scaling of wave function overlap~\cite{ChalkerDaniell} to the many-body case, and it is consistent with fractality of wave functions near MBLT~\cite{Luca13,Demler14,Santos15,Serbyn15}.

{\it Plasma model for level correlations.}
In the random matrix theory, the joint probability density for random matrix ensembles reads
\begin{eqnarray} \label{Eq:P-joint}
  P(\{s_i\})
  =
  \frac{e^{-\beta H}}{Z},
\
  H
  =
  \sum_i W(s_i)+\sum_{i<j} U(s_i-s_j),
\end{eqnarray}
where $\beta=1$ for orthogonal matrix ensemble which will be of primary interest. The confining potential $W(s) = s^2/2$ is parabolic, and interaction is $U(s_i-s_j) = -\ln |s_i-s_j|$. As Dyson demonstrated in his pioneering work~\cite{Dyson62}, this distribution function may be viewed as a stationary distribution of the stochastic random walk in a space of matrices (Hamiltonians).

To derive the joint distribution of eigenenergies from a random walk, one can start from the eigenbasis and perform a stochastic step in the space of Hamiltonians, induced by $\Delta H$. Then, we get the energy correction in a form
\begin{equation} \label{Eq:2opt}
\Delta s_n = V_{nn}+ \sum_{m\neq n} \frac{V_{mn} V_{nm}}{s_n-s_m},
\quad 
V_{mn} =\langle m|\Delta H  |n\rangle,
\end{equation}
which is the shift of eigenenergies induced by the perturbation $\Delta H$ up to second order. For Gaussian ensembles of random matrices, using $\langle V_{nm} V_{mn}\rangle = \frac2\beta \Delta \tau$ and $\langle V_{nn} V_{mm}\rangle = \delta_{mn}\Delta \tau$ one can derive Fokker-Planck equation (see Supplemental Material~\cite{suppmat} for more details). Its stationary (equilibrium) solution  is given by Eq.~(\ref{Eq:P-joint}) with logarithmic interaction.

Dyson's mapping was generalized to the case of disordered problems~\cite{ChalkerRW}. For such problems,  it is natural to perform a random walk (RW) in a space of Hamiltonians by changing realizations of disorder. As we are going to concentrate on properties of a spin chain in a random magnetic field, which is coupled to the $z$ component of a spin $S^z_i$, we take $\Delta H = \sum_{i=1}^L h_i(\tau) S^z_i$, with  $\langle h_i(\tau) h_j(\tau') \rangle = v^2 \delta(\tau-\tau')\delta_{ij}$. Similar to the case of random matrices~\cite{Dyson62,HaakeBook,suppmat}, the two correlators which determine the level dynamics are:
\begin{eqnarray} \label{Eq:Vnn-MBL}
  \langle V_{nn}V_{mm}\rangle  &=& 
  \delta d_{nm} =  \langle n|S^z_i|n\rangle  \langle m|S^z_i|m\rangle,\\
  \label{Eq:Vnm-MBL}
  \langle V_{nm}V_{mn}\rangle  &=& 
  \delta c_{nm} =  |\langle m|S^z_i|n\rangle|^2,
\end{eqnarray}
where we assumed that $v^2=\delta/L$, where $\delta$ is the many-body level spacing, so that $s_n$ represent  unfolded energy spectrum.  The correlator~(\ref{Eq:Vnn-MBL}) sets the spectrum of a random noise, while spectral function $c_{nm}$ determines the interaction between levels in the ensemble. 

{\it Effective interaction between levels.}  The RW process depends crucially on two correlators~Eqs.~(\ref{Eq:Vnn-MBL})-(\ref{Eq:Vnm-MBL}). To make analytic progress we use a mean-field like approximation~\cite{ChalkerRW}, assuming that $d_{nm}$ and $c_{nm}$ can be replaced by their ensemble averages,
\begin{equation} \label{Eq:cnm}
  c(\omega) = \langle c_{nm}\delta(s_n-s_m-\omega)\rangle,
\end{equation}
(and similar expression for $d_{nm}$) which now depend only on the energy difference between eigenstates. For the single-particle Anderson localization, the $c_{nm}$ and $d_{nm}$ necessarily coincide with the wave functions overlaps~\cite{ChalkerRW}, $c_{nm}=d_{nm} \propto \int dx  |\psi_{n}(\tau,x)|^2|\psi_{m}(\tau,x)|^2$. The fractality of the wave function near the mobility edge results in a power-law enhancement of~$c(\omega)\propto A/\omega^\gamma$~\cite{ChalkerDaniell,KravtsovEigenfunctionCorrelation}. 
In the case of ALT this enhancement  arises because the envelope of wave functions nearby in energy lives on the same multifractal domain~\cite{KravtsovEigenfunctionCorrelation}. In the many-body case  similar enhancement can arise from the fractal structure of the wave function in the Hilbert space in a vicinity of MBLT~\cite{Luca13,Demler14,Santos15,Serbyn15}. 

Inspired by the approach recently proposed in Ref.~\cite{Serbyn15}, we apply the fractal scaling to the matrix elements of a local operators. In particular, we assume that the inverse participation ratio (IPR), $I_2  = {\cal V}\sum_j |V_{ij}|^4  \propto {\cal V}^{-d_2}$, where $d_2$ is generalized fractal dimension, and ${\cal V}=\exp(sL)$ is the number of states in the Hilbert space. Using scaling, we translate the IPR into the scaling with the distance in the Hilbert space as ${\cal V}^2 \langle V_{ii}^2V_{ik}^2\rangle \propto ({\cal V}/{\cal R})^{1-d_2}$, where ${\cal R} = \exp(s d_{i,k})$ grows exponentially with (humming) distance in the Hilbert space, $d_{i,k}$. From here, expressing  $\cal V$ via frequency, as $\delta_{\cal V} = J/{\cal{V}} = \omega$, we  get: ${\cal V}^2 \langle V_{ii}^2V_{ki}^2\delta(E_i-E_k-\omega)\rangle \propto (J/\omega)^{1-d_2}$. Finally, omitting the diagonal matrix element, we arrive to the scaling:
\begin{equation} \label{Eq:c-metal}
  c(\omega) \propto \left(\frac{J}{\omega}\right)^{\gamma}, 
  \qquad
  \gamma = 1-d_2.
\end{equation}
Note, that we did not discuss the microscopic nature of a fractal behavior, although Griffiths (rare-region) effects~\cite{Demler14} in vicinity of MBL transition is one possible microscopic scenario. Also, relating $d_2$ to the properties of matrix elements, i.e.  exponent $\kappa$ in the scaling~\cite{Rahul14,Serbyn15}, $|V_{nm}| \propto \exp(-(s+\kappa) L)$ is an interesting question. 

The correlation between diagonal matrix elements, the function $d_{nm}$ also shows a power-law dependence. However, there is an enhancement of $d_{nm}$ for $n=m$, allowing to approximate $d(\omega)$ as a delta-function, see SM for additional discussion~\cite{suppmat}. 

{\it Implications for spectral statistics.} Using power-law form of $c(\omega)$~Eq.~(\ref{Eq:c-metal}), and the delta-function form of $d(\omega)$ we can map our model onto the plasma model for the level statistics~\cite{KravtsovPlasma}, provided $\gamma<1$. The plasma model assumes a power-law interaction potential  $U(s) = A/|s|^\gamma$ in the joint distribution function~(\ref{Eq:P-joint}). It  predicts the tails of the level statistics $P(s)\propto s^\beta \exp(-h_\gamma s^{2-\gamma})$ for  $s\gg1$, and variance of the number of levels in a box of size $N$ becomes $\mathop{\rm var} N \propto N^\gamma$, which is intermediate between WD-like rigidity $\mathop{\rm var} N \propto \log N$ and Poisson case~\cite{HaakeBook,MehtaBook}. 

For larger values of $\gamma\geq 1$ the effective interaction in the gas of eigenvalues becomes short range, and mapping to the plasma model no longer works. Instead, spectral properties now are expected to be well-described by a family of semi-Poisson distributions~\cite{Bogomolny99}, which arise from a gas of eigenvalues with a finite-range interaction. They predict Poisson-like behavior of the tails of $P(s)$ and level compressibility  $P(s) \propto s^\beta e^{-(\beta h+1)s}$, and  $\mathop{\rm var} N \propto  \chi N$ with $\chi \leq 1$, where $h$ is the range of interactions. Such level statistics has been dubbed ``critical'' in the literature~\cite{KravtsovNewClass,MuttalibNewFamily,VerbaarchotCritical} and is believed to describe the level statistics at the ALT~\cite{MirlinRMP,MirlinRPP}.

\begin{figure}[b]
\begin{center}
\includegraphics[width=0.99\columnwidth]{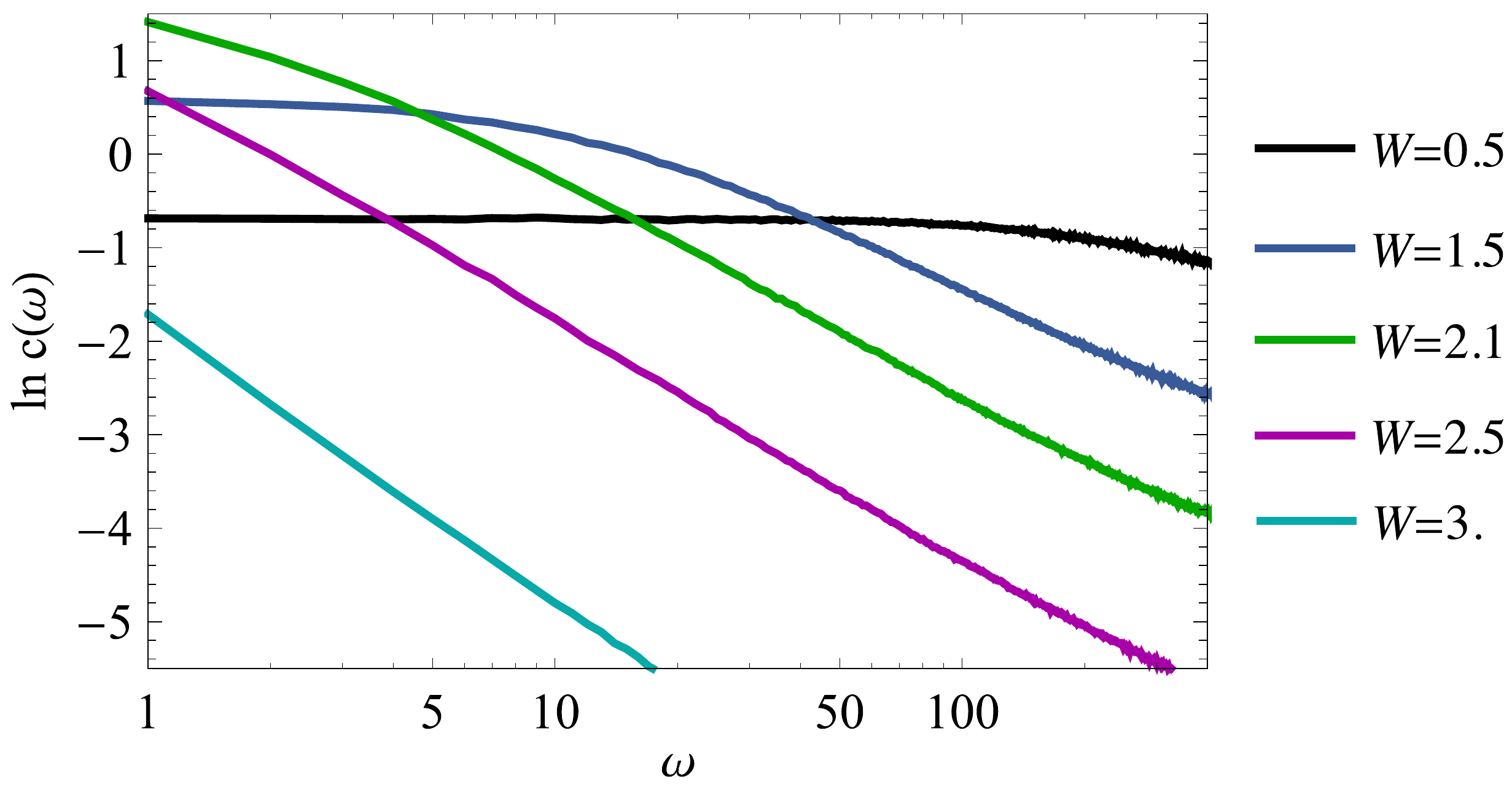}
\caption{ \label{Fig:cmn}  Averaged function $c(E)$ evolves from being almost flat at low disorder ($W=0.5$) to a power-law decay. Note that for the intermediate values of disorder, matrix element is enhanced at small energy difference compared to the limit of weak disorder.  
}
\end{center}
\end{figure}

Using the above intuition, we propose the following form of the level spacing distribution and spectral rigidity to interpolate between WDS and PS, 
\begin{equation} \label{Eq:P-interm}
  P(s;\beta, \gamma_P)
  = 
C_1  x^\beta \exp\left(- C_2 x^{2-\gamma_P}\right),
\
\mathop{\rm var} N  = \chi N^{\gamma_{\rm var}},
\end{equation}
where the parameter $1\geq\gamma_P,\gamma_{\rm var}\geq0$ controls the tails of the statistics and level rigidity, and $1\geq\beta\geq0$ determines the level repulsion.  The constants $C_{1,2}$ can be fixed by requiring that $\langle 1\rangle=\langle s\rangle=1$.  When $\gamma_P=0$, this distribution becomes WD. In the opposite limit, $\gamma_P\to 1$, distribution~(\ref{Eq:P-interm}) becomes a semi-Poission with generic $\beta$. For the spectral rigidity our interpolating function also can describe the (semi-)Poisson limit, however failing to capture logarithmic growth of $\mathop{\rm var} N$ in the WD case.

\begin{figure*}
\begin{center}
\includegraphics[width=0.659\columnwidth]{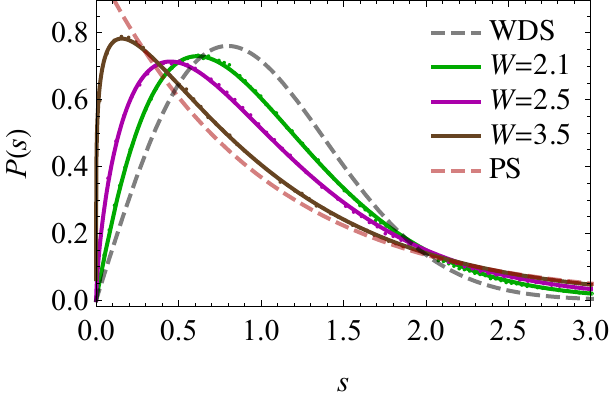}
\includegraphics[width=0.659\columnwidth]{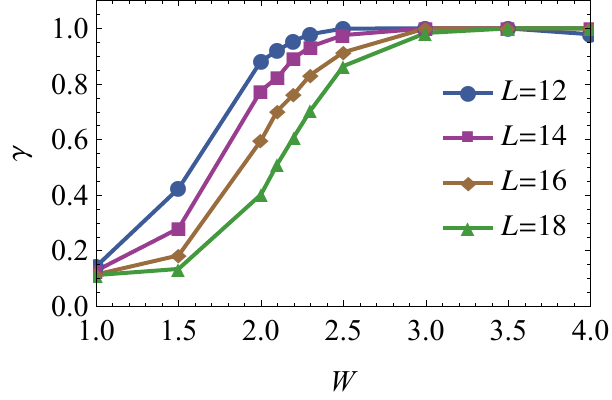}
\includegraphics[width=0.659\columnwidth]{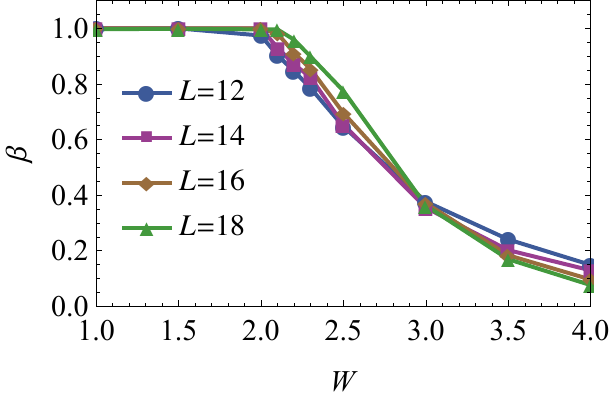}\\
\caption{ \label{Fig:gammabeta} 
(a) Evolution of level spacing distributions as system is tuned towards MBL phase. Points represent data, while solid lines are best fits with a two-parameter distribution~(\ref{Eq:P-interm}). Red and black dashed lines correspond  to Poisson and Wigner-Dyson distribution.
(b) The exponent $\gamma_P$, controlling tails of level statistics, flows with $L$ for $W\lesssim 2.5$, but is constant in vicinity of MBLT $W_c\approx 3.6$. (c) In contrast, $\beta$ controlling the level repulsion, remains constant for $W\lesssim 2$, and starts to flow closer to the MBLT.}
\end{center}
\end{figure*}

{\it Numerical results.} We use the XXZ spin chain in a random field as a specific model with a previously located MBL transition~\cite{PalHuse} to test our picture of level statistics. The Hamiltonian is\begin{equation} \label{Eq:XXZ}
\hat H_\text{XXZ} = \sum_{\langle ij \rangle} {\vec S}_i\cdot {\vec S}_j+  \sum_i w_i  S_i^z,
\quad  S^{x,y,z}  = \frac12 \vec \sigma^{x,y,z},
\end{equation}
where  disorder enters via random fields $w_i$ uniformly distributed in the interval $ [-W;W]$.  We perform exact diagonalization for chains of size $L=12,\ldots 16 (18)$  with periodic boundary conditions to extract properties of matrix elements (spectral statistics). We use central part of the many-body spectrum, which corresponds to energy density $\epsilon = (E-E_\text{min})/(E_\text{max}-E_\text{min})= 0.45\pm 0.1$. The MBL transition at this energy density is believed to occur near $W_c\approx 3.6$~\cite{Alet14}. To unfold levels, we fit the staircase function with a 3rd order polynomial. We use both local and global level unfolding schemes~\cite{Misleading}. 

We start by discussing the numerical results for averaged $c(\omega)$, presented in Fig.~\ref{Fig:cmn}(a). Upon increasing disorder, we see the crossover of $c(\omega)$ from a constant to a power-law decay. As one may expect, this crossover happens at some scale, $N_\text{erg}$, so that $c(\omega<N_\text{erg})\propto \text{const}$, and decays as a power-law beyond  $\omega>N_\text{erg}$. The additional scale $N_\text{erg}$ has a meaning similar to the correlation length, over which ergodicity holds. As $N_\text{erg}\to 0$, interaction between levels becomes critical even for the smallest separations.

From the power-law form of $c(\omega)$, we expect that  level spacing distribution for the XXZ spin chain to be well described by Eq.~(\ref{Eq:P-interm}). Fig.~\ref{Fig:gammabeta}(a) illustrates that the flow of the level statistics is indeed well captured by Eq.~(\ref{Eq:P-interm}). The $P(s)$ for disorder $W<2$ is not shown, as it looks very similar to WD distribution: since $P(s)$ is influenced the most by the interaction between close levels,  $N_\text{erg}$ must become close to zero before we see the flow in the level statistics. In contrast to the level statistics, which is influenced by a non-critical part of $c(\omega)$, the spectral rigidity is expected to be less sensitive to the behavior of $c(\omega)$ at small $\omega$. In SM~\cite{suppmat}  we  show that $\mathop{\rm var} N$  behaves as a power-law~(\ref{Eq:P-interm}), and becomes linear for~$W\gtrsim 2$. Also, we test that different estimates for exponent $\gamma$ show reasonable agreement as follows from plasma model.

Finally, we consider the flow of parameters $\gamma_P$ and $\beta$ with increasing system size, presented in Fig.~\ref{Fig:gammabeta}(b)-(c). While $\gamma_P$ controlling tails of the level statistics has a strong flow at disorder $W\leq 2.5$, at larger disorders $\gamma_P$ is very close to one and changes little with $L$. This further supports the conclusion that for $W\geq 2.5$ the effective interaction between energy levels becomes short-ranged for the largest accessible system sizes.  Consistent with our expectation, $\beta$ shown in  Fig.~\ref{Fig:gammabeta}(c) changes weakly when statistics is described by plasma model ($W\leq2$), and begins to flow once level interactions are local. 

{\it Discussion and open questions.} Using analytical and numerical arguments we described the spectral properties across MBL transition using a two-stage flow picture. Note that we need at least two parameters, $\gamma$ and $N_\text{erg}$, to describe level statistics. This is not surprising if we recall that even the case of ALT, the existence of multifractality means that to describe the universal properties one requires more information beyond the small number of critical indices needed for a simple thermodynamic phase transition~\cite{MirlinRMP,MirlinRPP}.  Below we discuss the implications of the proposed picture of the spectral statistics flow.

At the first stage the ``correlation length'' $N_\text{erg}$ shrinks to zero, but the exponent responsible for level interactions $\gamma$ is smaller than one.  Intuitively, the levels beyond correlation length become more and more different, corresponding to a gradual breakdown of the ETH. Here the level statistics can be described by the effective plasma model. Although this model was proposed some time ago~\cite{KravtsovPlasma}, it does not apply in the case of ALT, despite the presence of multifractality near single particle mobility edge. Hence, to the best of our knowledge, the present study is the first physical realization of the plasma model.  

The second stage begins at $W\geq 2.5$, when $\gamma \geq 1$ so that interactions between levels are local. Although we cannot exclude the finite size effects, the numerical estimates for the MBL transition at $W_c\approx 3.6$ suggest that \emph{at the MBL transition} interactions between levels are local. Thus, we conjecture that level statistics near and at the MBLT belongs to the same or similar ``critical'' family as the universal statistics at the ALT~\cite{KravtsovNewClass,MuttalibNewFamily,VerbaarchotCritical}.  This also naturally explains why the average ratio of the level spacing $r=\min(\delta_{n},\delta_{n+1})/\max(\delta_{n},\delta_{n+1})$  at the MBLT, widely used in the literature~\cite{PalHuse,Alet14,Demler14,Reichman14}, is very close to the value expected from PS.  

The semi-Poisson level statistics emerges at the same value of disorder where the boundary of the Griffiths phase was previously identified in the literature~\cite{Demler14}, $W\approx 2.5$~(Refs.~\cite{Luca13,Goold15} report the onset of ergodicity breaking at the same location). The existing theories of the MBLT~\cite{AltmanRG14,Potter15} predict extensive entanglement and subdiffusive transport in the ergodic phase. The wide region of critical statistics near transition may be a manifestation of finite size effects  (system sizes studied are smaller that diverging correlation length). Indeed the strong overlaps only between adjacent energy levels imply logarithmic transport~\cite{Serbyn15}, predicted at the MBLT~\cite{AltmanRG14,Potter15}.  On the other hand,  existence of thermodynamically stable Griffiths phase is another intriguing possibility. 

In closing, we have found that Dyson's mapping of level statistics to Brownian motion allows one to understand the spectral statistics in the MBL transition at least as well as in the ALT for which it was introduced. There are basic differences between the two transitions, e.g., several quantities which are uniquely defined at the ALT allow inequivalent generalizations to the MBLT.  There are two steps of the spectral statistics flow, one with long-range interactions (the plasma model) and one with local interactions, and the boundary between the two is found numerically to coincide with the onset of a Griffiths phase and subdiffusive transport.  Since level statistics are known to be the simplest universal probe of the transition to quantum chaos in simpler problems, understanding the origin and universality of the two-step plasma model of level statistics is an important challenge for theories of the MBLT.
 
{\it Acknowledgements.} 
M.S. acknowledges useful discussions with V. Kravtsov, D. Abanin, Z. Papic, E. Mucciolo, A.C. Potter, R. Vasseur, and S. Gazit. M.S. was supported by Gordon and Betty Moore Foundation's EPiQS Initiative through Grant GBMF4307.  J.E.M. was supported by NSF DMR-1206515 and the Simons Foundation.

\setcounter{figure}{0}
\makeatletter 
\renewcommand{\thefigure}{S\@arabic\c@figure} 

\setcounter{equation}{0}
\makeatletter 
\renewcommand{\theequation}{S\@arabic\c@equation}

\clearpage
\newpage

\onecolumngrid
\appendix

\begin{center}
{\large \bf Supplemental Online Material for ``Spectral statistics across the many-body localization transition"}
\end{center}
\vspace{10pt}
\twocolumngrid

Below we  present additional details on the derivation of the level statistics from the Brownian motion. In particular, we discuss the approximation for $d(\omega)$ used in the main text. In the second part we discuss the behavior of the spectral rigidity, and compare various estimates for exponent $\gamma$.

\subsection{Analytic derivation of level statistics from the Brownian motion}
\subsubsection{Wigner-Dyson statistics}
Let us begin with reproducing the WD statistics from the Brownian motion model. We assume the Brownian motion in the space of random matrices, 
\begin{equation} \label{Eq:dH}
 H(\tau) = H_0+\int_0^\tau d\tau\, V(\tau),
\end{equation}
where matrix $V(\tau)$ satisfies the following properties:
\begin{eqnarray} \label{Eq:Random-diag}
\langle V_{nm}(\tau) \rangle &=& -H_{nm}(\tau),\\
\langle V_{nm}(\tau) V_{mn}(\tau') \rangle &=&  \delta(\tau-\tau'), \\ \label{Eq:Random-offdiag}
\langle V_{nn}(\tau) V_{mm}(\tau') \rangle &=&  \frac{2}{\beta_{RM}} \delta(\tau-\tau')\delta_{nm}.
\end{eqnarray}
Parameter $\beta_{RM}$ specifies the symmetry class, $\beta_{RM}=1$ for GOE, and $\beta_{RM}=2$ for GUE. 

For convenience, we fix the basis to coincide with the (instantaneous) eigenbasis of $H(\tau)$. We  apply the perturbation theory to calculate correction to eigenvalues of $H(\tau)$, $\{s_n\}$ induced by the change in the matrix
\begin{equation} \label{Eq:dV}
\Delta V  = \int_\tau^{\tau+\Delta \tau} d\tau \, V(\tau) .
\end{equation}
Resulting correction to $s_n$ reads:
\begin{equation} \label{Eq:sn-2opt}
  \Delta s_n = \Delta V_{nn}+\sum_{m\neq n} \frac{|\Delta V_{nm}|^2}{s_n-s_m}.
\end{equation}
Averaging this equation over $V$ using the fact that $H_{nn} = s_n$ in the eigenbasis, we get:
\begin{eqnarray} \label{Eq:ds}
  \langle \Delta s_n \rangle
  &=&
  \mu_n(\{s\})\Delta \tau,
  \\\label{Eq:drift}
    \mu_n[\{s\}]
    &=&
    - s_n + \sum_{m\neq n} \frac{1}{s_n-s_m}.
\end{eqnarray} 
Using the second moment 
\begin{equation} \label{Eq:Dl2}
  \langle \Delta s_n \Delta s_m \rangle
  =
 \frac{2}{\beta_{RM}}\delta_{nm} \Delta \tau,
\end{equation}
we find that $s_n(\tau)$ obeys the following Langevin equation:
\begin{equation} \label{Eq:Langevin}
  \frac{d s_n}{d \tau}
  =
  \mu_n[\{s\}]
  + \xi_n (\tau),
\end{equation}
where drift term is given by Eq.~(\ref{Eq:drift}) and white noise is specified by
\begin{eqnarray} \label{Eq:noise}
  \langle \xi_{n}(\tau )\xi_m (\tau')\rangle
  &=&
 \frac{2}{\beta_{RM}} \delta_{nm}\delta(\tau-\tau').
\end{eqnarray}

Translating the Langevin equation into the Fokker-Planck  equation governing the evolution of the joint probability distribution function $P(\{s_n\},\tau)$, we get:
\begin{multline} \label{Eq:FP}
\frac{\partial P(\{s\},\tau)}{\partial \tau}
=
\sum_n \left\{
-\frac{\partial }{\partial s_n} [ \mu_n(\{s\}) P(\{s\},\tau)] \right.
\\ \left.
+ \frac{1}{\beta_{RM}} \frac{\partial^2  }{\partial s_n^2}  P(\{s\},\tau)]
\right\}.
\end{multline}
Using explicit expression for the drift, Eq.~(\ref{Eq:drift}) we see that the joint probability distribution
\begin{equation} \label{Eq:PWD}
P(\{s\})
=
C \prod_{m<n} |s_n-s_m|^{\beta_{RM}} \exp\left(-\frac{\beta_{RM}}{2} \sum_n s_n^2\right)
\end{equation}
is a stationary solution of the Fokker-Planck equation~(\ref{Eq:FP}). This distribution function corresponds to Wigner-Dyson statistics. It is equivalent to the partition function of plasma with logarithmic repulsion and a parabolic confining potential. 
\begin{figure*}
\begin{center}
\includegraphics[width=1.99\columnwidth]{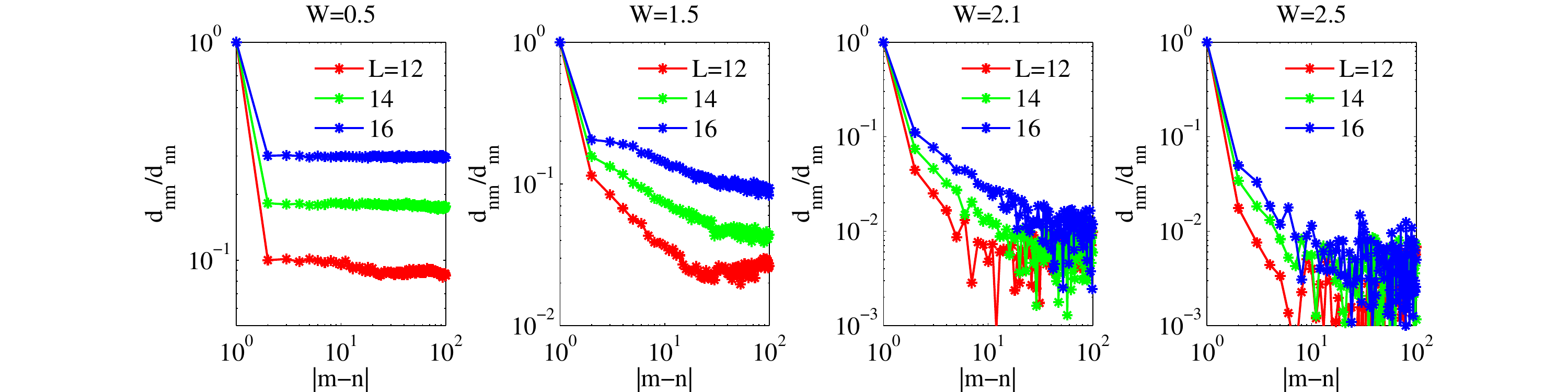}\\
\caption{ \label{Fig:dnm} Normalized diagonal correlator $\langle d_{nm}\rangle /\langle d_{nn}\rangle$ as a function of $|m-n|$ for different values of disorder. Although for intermediate disorders $\langle d_{nm}\rangle$ decays as a power-law with $|n-m|$, the value of $d_{nn}$ is significantly enhanced, compared to the power-law tail. }
\end{center}
\end{figure*}
\subsubsection{Plasma model}
After discussing the random matrix example, we move on to the many-body case. There the Brownian motion is induced by a random walk over different realizations of disorder. For a specific case of XXZ spin chain in a random magnetic field, we have
\begin{equation} \label{Eq:Vt}
  V(\tau) = \sum_{i=1}^L h_i(\tau)S^z_i,
  \quad
  \langle h_i(\tau) h_j(\tau)\rangle = v^2\delta(\tau-\tau')\delta_{ij}.
\end{equation}
Repeating the procedure outlined above for the case of random matrices, we get the following Langevin equation:
\begin{equation} \label{Eq:LangevinMB}
  \frac{d s_n}{d \tau}
  =
\mu_n[\{s\},\tau]
  +  \xi_n (\tau),
\end{equation}
with drift and noise terms reading:
\begin{eqnarray} \label{Eq:driftMB}
 \mu_n[\{s\},\tau]
    &=&
   \frac{v^2}{\delta}\sum_{i=1}^L \sum_{m\neq n} \frac{c^i_{nm}(\tau)}{s_n-s_m},
  \\   \label{Eq:noiseMB}
  \langle  \xi_{n}(\tau )  \xi_m (\tau')\rangle
  &=&
   \frac{v^2}{\delta} \sum_{i=1}^L d^i_{nm}(\tau)\delta(\tau-\tau'),
\end{eqnarray} 
where the level spacing $\delta$ appeared since we assume that $s$ represents unfolded energy spectrum with mean level spacing of one. 
The correlation functions $c_{nm}$ and $d_{nm}$ are defined by the matrix elements of the operator $S^z_i$ on a site $i$ (which couples to the random magnetic field) in the instantaneous eigenbasis:
\begin{eqnarray} \label{Eq:c}
 c^i_{nm}(\tau) &=& \frac{1}{\delta}\langle n |S^z_i | m \rangle \langle m |S^z_i | n \rangle,
  \\ \label{Eq:d}
 d^i_{nm}(\tau) &=& \frac{1}{\delta}\langle n |S^z_i | n \rangle \langle m |S^z_i | m \rangle.
\end{eqnarray}
The mean field approximation employed in the main text amounts to replacing $c^i_{nm}(\tau)$ and $d^i_{nm}(\tau)$ with their average over ensemble, which becomes only a function of $n-m$. Fixing $v^2=\delta/L$ to cancel the level spacing from resulting equations, we get the following Lagnevin dynamics:
\begin{equation} \label{Eq:LangevinMF}
  \frac{d s_n}{d \tau}
  =
 \mu_n[\{s\},\tau]
  + \xi_n (\tau),
\end{equation}
with  drift and noise terms
\begin{eqnarray} \label{Eq:driftMF}
    \mu_n[\{s\},\tau]
    &=&
    \sum_{m\neq n} \frac{c(s_n-s_m)}{s_n-s_m},
  \\   \label{Eq:noiseMF}
  \langle \xi_{n}(\tau ) \xi_m (\tau')\rangle
  &=&
   d(s_n-s_m)\delta(\tau-\tau'), 
\end{eqnarray} 
 expressed via rescaled correlators:

Assuming a delta-function form of $d(\omega)$, and a power-law ansatz for $c(\omega)$,
\begin{eqnarray} \label{Eq:cw}
 c(\omega) = \frac{A}{|\omega|^\gamma}, 
 \quad 
0<\gamma<1,
\end{eqnarray}
the stationary solution of corresponding Fokker-Planck equation reproduces the partition function of plasma with interaction potential given by:
\begin{eqnarray} \label{Eq:Up}
  U(s) = \frac{C}{|s|^\gamma}.
\end{eqnarray}

\subsubsection{Numerical results for $d(\omega)$}
In order to check the validity of approximating $d(\omega)$ by a delta-function, we calculate $d_{nm}$ numerically for the XXZ spin chain. Figure~\ref{Fig:dnm} reveals the evolution of average $\langle d_{nm}\rangle$ for different system sizes and disorders. At disorder $W=0.5$, when system is in the ergodic phase, we see that $d_{nm}$ when $|m-n|>0$ does not depend on $|m-n|$, which is a manifestation of the eigenstate thermalization  hypothesis(ETH).  Nevertheless, values of $\langle d_{nm}\rangle$ for $n\neq m$ are suppressed compared to $d_{nn}$. Even nearby eigenstates can have average spin of different sign, this does not contradict to the ETH which rigorously applies to coarse-grained observables. 

For larger disorder, $W>1.5$, the $d_{nm}$ decays approximately as power-law. Nevertheless, the value of $d_{nn}$ is still considerably enhanced, hence a proper approximation for the $d(\omega)$ is $d(\omega) = c_1 \delta(\omega)+ c_2/\omega^{\gamma'}$ [scaling of the relative weight of delta function in $d(\omega)$ in the thermodynamic limit is an interesting and open question]. Presence of the delta-function contribution in $d(\omega)$ is sufficient to make the  spectral statistics different compared to the  case of Anderson transition. In particular, repeating the mean-field treatment of Ref.~\cite{ChalkerRW}, we do not get constant spectral form-factor $K(t,\tau)$ in the limit $t\to 0$. This suggests that level compressibility is vanishing, consistent with plasma model~\cite{KravtsovPlasma}.

\subsection{Spectral rigidity and different estimates for $\gamma$ across the MBL transition}
\begin{figure}[h]
\begin{center}
\includegraphics[width=0.9\columnwidth]{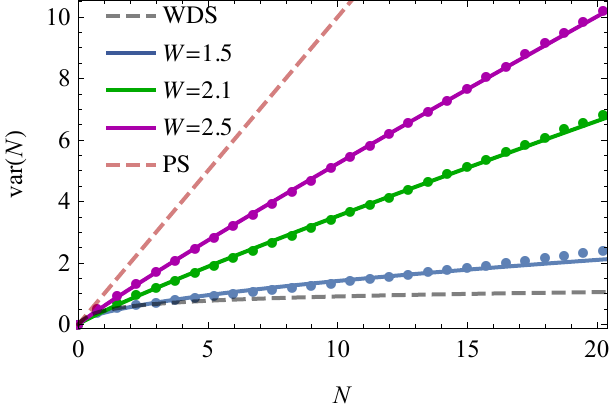}
\caption{ \label{Fig:varN} Variance of the level number in the box of size $N$ becomes essentially linear in $N$ for disorder above $W\geq 2.5$. To minimize the influence of the unfolding procedure, we show and use the part where results agree for both local and global unfolding. All data is for $L=16$ spins.
}
\end{center}
\end{figure}

To probe the spectral rigidity, we study the behavior of the variance of number of levels in the box of size $N$. The variance  $\mathop{\mathrm{var}} N$ as a function of the box size is shown in Fig.~\ref{Fig:varN} for $L=16$ spins, along with the power-law fits of its behavior. The exponent extracted from the fits, $\gamma_{\text{var}}$ below is compared to exponents extracted by other means.

\begin{figure}[h]
\includegraphics[width=0.9\columnwidth]{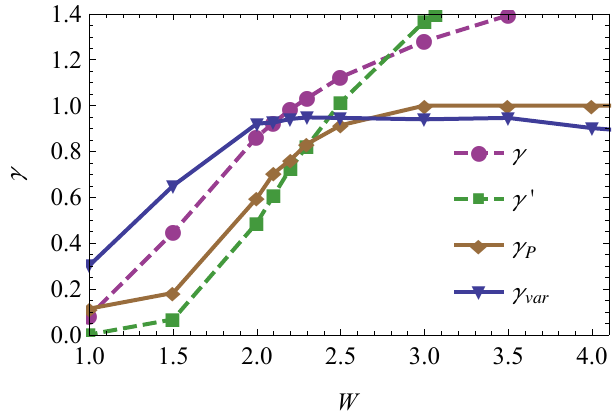}\\
\caption{ \label{Fig:gamma-app} Exponent $\gamma$ extracted from $c(\omega)$ agrees reasonably with the exponent $\gamma_\text{var}$ from the tails of the level statistic. The exponent $\gamma_P$ governing the tails of the $P(s)$ is consistently smaller, but agrees with $\gamma'= (d \ln c(\omega)/d\ln \omega)_{\omega=0}$.}
\end{figure}

Three exponents, $\gamma, \gamma_P, \gamma_\text{var}$, obtained from matrix elements, level statistics, and level rigidity respectively,  are expected to coincide in the region of applicability of plasma model. In Fig.~\ref{Fig:gamma-app} we plot these exponents as a function of disorder for spin chains with $L=16$ spins.  We observe the reasonable agreement between the values of $\gamma_\text{var}$ and $\gamma$ measured from the tails of level statistics. On the other hand, $\gamma_P$ is consistently smaller, and matches much better the $\gamma'$, logarithmic derivative of $c(\omega)$ at $\omega\to 0$, confirming that $P(s)$ is more strongly influenced by a non-universal part in the $c(\omega)$. Both $\gamma$ and $\gamma'$ become larger than one at $W\approx 2.5$, indicating that for larger disorders the level statistics enters a semi-Poisson regime.


\begin{thebibliography}{42}%
\makeatletter
\providecommand \@ifxundefined [1]{%
 \@ifx{#1\undefined}
}%
\providecommand \@ifnum [1]{%
 \ifnum #1\expandafter \@firstoftwo
 \else \expandafter \@secondoftwo
 \fi
}%
\providecommand \@ifx [1]{%
 \ifx #1\expandafter \@firstoftwo
 \else \expandafter \@secondoftwo
 \fi
}%
\providecommand \natexlab [1]{#1}%
\providecommand \enquote  [1]{``#1''}%
\providecommand \bibnamefont  [1]{#1}%
\providecommand \bibfnamefont [1]{#1}%
\providecommand \citenamefont [1]{#1}%
\providecommand \href@noop [0]{\@secondoftwo}%
\providecommand \href [0]{\begingroup \@sanitize@url \@href}%
\providecommand \@href[1]{\@@startlink{#1}\@@href}%
\providecommand \@@href[1]{\endgroup#1\@@endlink}%
\providecommand \@sanitize@url [0]{\catcode `\\12\catcode `\$12\catcode
  `\&12\catcode `\#12\catcode `\^12\catcode `\_12\catcode `\%12\relax}%
\providecommand \@@startlink[1]{}%
\providecommand \@@endlink[0]{}%
\providecommand \url  [0]{\begingroup\@sanitize@url \@url }%
\providecommand \@url [1]{\endgroup\@href {#1}{\urlprefix }}%
\providecommand \urlprefix  [0]{URL }%
\providecommand \Eprint [0]{\href }%
\providecommand \doibase [0]{http://dx.doi.org/}%
\providecommand \selectlanguage [0]{\@gobble}%
\providecommand \bibinfo  [0]{\@secondoftwo}%
\providecommand \bibfield  [0]{\@secondoftwo}%
\providecommand \translation [1]{[#1]}%
\providecommand \BibitemOpen [0]{}%
\providecommand \bibitemStop [0]{}%
\providecommand \bibitemNoStop [0]{.\EOS\space}%
\providecommand \EOS [0]{\spacefactor3000\relax}%
\providecommand \BibitemShut  [1]{\csname bibitem#1\endcsname}%
\let\auto@bib@innerbib\@empty
\bibitem [{\citenamefont {Srednicki}(1994)}]{SrednickiETH}%
  \BibitemOpen
  \bibfield  {author} {\bibinfo {author} {\bibfnamefont {M.}~\bibnamefont
  {Srednicki}},\ }\bibfield  {title} {\emph {\bibinfo {title} {Chaos and
  quantum thermalization},\ }}\href {\doibase 10.1103/PhysRevE.50.888}
  {\bibfield  {journal} {\bibinfo  {journal} {Phys. Rev. E}\ }\textbf {\bibinfo
  {volume} {50}},\ \bibinfo {pages} {888} (\bibinfo {year} {1994})}\BibitemShut
  {NoStop}%
\bibitem [{\citenamefont {Rigol}\ \emph {et~al.}(2008)\citenamefont {Rigol},
  \citenamefont {Dunjko},\ and\ \citenamefont {Olshanii}}]{RigolNature}%
  \BibitemOpen
  \bibfield  {author} {\bibinfo {author} {\bibfnamefont {M.}~\bibnamefont
  {Rigol}}, \bibinfo {author} {\bibfnamefont {V.}~\bibnamefont {Dunjko}}, \
  and\ \bibinfo {author} {\bibfnamefont {M.}~\bibnamefont {Olshanii}},\
  }\bibfield  {title} {\emph {\bibinfo {title} {Thermalization and its
  mechanism for generic isolated quantum systems},\ }}\href {\doibase
  10.1038/nature06838} {\bibfield  {journal} {\bibinfo  {journal} {Nature}\
  }\textbf {\bibinfo {volume} {452}},\ \bibinfo {pages} {854} (\bibinfo {year}
  {2008})}\BibitemShut {NoStop}%
\bibitem [{\citenamefont {Haake}(2013)}]{HaakeBook}%
  \BibitemOpen
  \bibfield  {author} {\bibinfo {author} {\bibfnamefont {F.}~\bibnamefont
  {Haake}},\ }\href {https://books.google.com/books?id=Glv9CAAAQBAJ} {\emph
  {\bibinfo {title} {Quantum Signatures of Chaos}}},\ Springer Series in
  Synergetics\ (\bibinfo  {publisher} {Springer Berlin Heidelberg},\ \bibinfo
  {year} {2013})\BibitemShut {NoStop}%
\bibitem [{\citenamefont {Mehta}(2014)}]{MehtaBook}%
  \BibitemOpen
  \bibfield  {author} {\bibinfo {author} {\bibfnamefont {M.}~\bibnamefont
  {Mehta}},\ }\href {https://books.google.com/books?id=\_MjSBQAAQBAJ} {\emph
  {\bibinfo {title} {Random Matrices: Revised and Enlarged Second Edition}}},\
  Pure and Applied Mathematics\ (\bibinfo  {publisher} {Elsevier Science},\
  \bibinfo {year} {2014})\BibitemShut {NoStop}%
\bibitem [{\citenamefont {Basko}\ \emph {et~al.}(2006)\citenamefont {Basko},
  \citenamefont {Aleiner},\ and\ \citenamefont {Altshuler}}]{Basko06}%
  \BibitemOpen
  \bibfield  {author} {\bibinfo {author} {\bibfnamefont {D.}~\bibnamefont
  {Basko}}, \bibinfo {author} {\bibfnamefont {I.}~\bibnamefont {Aleiner}}, \
  and\ \bibinfo {author} {\bibfnamefont {B.}~\bibnamefont {Altshuler}},\
  }\bibfield  {title} {\emph {\bibinfo {title} {Metal--insulator transition in
  a weakly interacting many-electron system with localized single-particle
  states},\ }}\href {\doibase http://dx.doi.org/10.1016/j.aop.2005.11.014}
  {\bibfield  {journal} {\bibinfo  {journal} {Annals of Physics}\ }\textbf
  {\bibinfo {volume} {321}},\ \bibinfo {pages} {1126 } (\bibinfo {year}
  {2006})}\BibitemShut {NoStop}%
\bibitem [{\citenamefont {Gornyi}\ \emph {et~al.}(2005)\citenamefont {Gornyi},
  \citenamefont {Mirlin},\ and\ \citenamefont {Polyakov}}]{Mirlin05}%
  \BibitemOpen
  \bibfield  {author} {\bibinfo {author} {\bibfnamefont {I.~V.}\ \bibnamefont
  {Gornyi}}, \bibinfo {author} {\bibfnamefont {A.~D.}\ \bibnamefont {Mirlin}},
  \ and\ \bibinfo {author} {\bibfnamefont {D.~G.}\ \bibnamefont {Polyakov}},\
  }\bibfield  {title} {\emph {\bibinfo {title} {Interacting electrons in
  disordered wires: Anderson localization and low-$t$ transport},\ }}\href
  {\doibase 10.1103/PhysRevLett.95.206603} {\bibfield  {journal} {\bibinfo
  {journal} {Phys. Rev. Lett.}\ }\textbf {\bibinfo {volume} {95}},\ \bibinfo
  {pages} {206603} (\bibinfo {year} {2005})}\BibitemShut {NoStop}%
\bibitem [{\citenamefont {Huse}\ and\ \citenamefont
  {Oganesyan}(2013)}]{Huse13}%
  \BibitemOpen
  \bibfield  {author} {\bibinfo {author} {\bibfnamefont {D.~A.}\ \bibnamefont
  {Huse}}\ and\ \bibinfo {author} {\bibfnamefont {V.}~\bibnamefont
  {Oganesyan}},\ }\bibfield  {title} {\emph {\bibinfo {title} {A phenomenology
  of certain many-body localized systems},\ }}\href@noop {} {\bibfield
  {journal} {\bibinfo  {journal} {ArXiv e-prints}\ } (\bibinfo {year}
  {2013})},\ \Eprint {http://arxiv.org/abs/arXiv:1305.4915} {arXiv:1305.4915}
  \BibitemShut {NoStop}%
\bibitem [{\citenamefont {Serbyn}\ \emph {et~al.}(2013)\citenamefont {Serbyn},
  \citenamefont {Papi\ifmmode~\acute{c}\else \'{c}\fi{}},\ and\ \citenamefont
  {Abanin}}]{Serbyn13-1}%
  \BibitemOpen
  \bibfield  {author} {\bibinfo {author} {\bibfnamefont {M.}~\bibnamefont
  {Serbyn}}, \bibinfo {author} {\bibfnamefont {Z.}~\bibnamefont
  {Papi\ifmmode~\acute{c}\else \'{c}\fi{}}}, \ and\ \bibinfo {author}
  {\bibfnamefont {D.~A.}\ \bibnamefont {Abanin}},\ }\bibfield  {title} {\emph
  {\bibinfo {title} {Local conservation laws and the structure of the many-body
  localized states},\ }}\href {\doibase 10.1103/PhysRevLett.111.127201}
  {\bibfield  {journal} {\bibinfo  {journal} {Phys. Rev. Lett.}\ }\textbf
  {\bibinfo {volume} {111}},\ \bibinfo {pages} {127201} (\bibinfo {year}
  {2013})}\BibitemShut {NoStop}%
\bibitem [{\citenamefont {Imbrie}(2014)}]{Imbrie14}%
  \BibitemOpen
  \bibfield  {author} {\bibinfo {author} {\bibfnamefont {J.~Z.}\ \bibnamefont
  {Imbrie}},\ }\href@noop {} {\bibfield  {journal} {\bibinfo  {journal} {Arxiv
  e-prints}\ } (\bibinfo {year} {2014})},\ \Eprint
  {http://arxiv.org/abs/1403.7837} {1403.7837} \BibitemShut {NoStop}%
\bibitem [{\citenamefont {Ros}\ \emph {et~al.}(2015)\citenamefont {Ros},
  \citenamefont {M{\"u}ller},\ and\ \citenamefont
  {Scardicchio}}]{ScardicchioLIOM}%
  \BibitemOpen
  \bibfield  {author} {\bibinfo {author} {\bibfnamefont {V.}~\bibnamefont
  {Ros}}, \bibinfo {author} {\bibfnamefont {M.}~\bibnamefont {M{\"u}ller}}, \
  and\ \bibinfo {author} {\bibfnamefont {A.}~\bibnamefont {Scardicchio}},\
  }\bibfield  {title} {\emph {\bibinfo {title} {Integrals of motion in the
  many-body localized phase},\ }}\href {\doibase
  http://dx.doi.org/10.1016/j.nuclphysb.2014.12.014} {\bibfield  {journal}
  {\bibinfo  {journal} {Nuclear Physics B}\ }\textbf {\bibinfo {volume}
  {891}},\ \bibinfo {pages} {420 } (\bibinfo {year} {2015})}\BibitemShut
  {NoStop}%
\bibitem [{\citenamefont {Pal}\ and\ \citenamefont {Huse}(2010)}]{PalHuse}%
  \BibitemOpen
  \bibfield  {author} {\bibinfo {author} {\bibfnamefont {A.}~\bibnamefont
  {Pal}}\ and\ \bibinfo {author} {\bibfnamefont {D.~A.}\ \bibnamefont {Huse}},\
  }\bibfield  {title} {\emph {\bibinfo {title} {Many-body localization phase
  transition},\ }}\href {\doibase 10.1103/PhysRevB.82.174411} {\bibfield
  {journal} {\bibinfo  {journal} {Phys. Rev. B}\ }\textbf {\bibinfo {volume}
  {82}},\ \bibinfo {pages} {174411} (\bibinfo {year} {2010})}\BibitemShut
  {NoStop}%
\bibitem [{\citenamefont {Bar~Lev}\ \emph {et~al.}(2015)\citenamefont
  {Bar~Lev}, \citenamefont {Cohen},\ and\ \citenamefont
  {Reichman}}]{Reichman14}%
  \BibitemOpen
  \bibfield  {author} {\bibinfo {author} {\bibfnamefont {Y.}~\bibnamefont
  {Bar~Lev}}, \bibinfo {author} {\bibfnamefont {G.}~\bibnamefont {Cohen}}, \
  and\ \bibinfo {author} {\bibfnamefont {D.~R.}\ \bibnamefont {Reichman}},\
  }\bibfield  {title} {\emph {\bibinfo {title} {Absence of diffusion in an
  interacting system of spinless fermions on a one-dimensional disordered
  lattice},\ }}\href {\doibase 10.1103/PhysRevLett.114.100601} {\bibfield
  {journal} {\bibinfo  {journal} {Phys. Rev. Lett.}\ }\textbf {\bibinfo
  {volume} {114}},\ \bibinfo {pages} {100601} (\bibinfo {year}
  {2015})}\BibitemShut {NoStop}%
\bibitem [{\citenamefont {Luitz}\ \emph {et~al.}(2015)\citenamefont {Luitz},
  \citenamefont {Laflorencie},\ and\ \citenamefont {Alet}}]{Alet14}%
  \BibitemOpen
  \bibfield  {author} {\bibinfo {author} {\bibfnamefont {D.~J.}\ \bibnamefont
  {Luitz}}, \bibinfo {author} {\bibfnamefont {N.}~\bibnamefont {Laflorencie}},
  \ and\ \bibinfo {author} {\bibfnamefont {F.}~\bibnamefont {Alet}},\
  }\bibfield  {title} {\emph {\bibinfo {title} {Many-body localization edge in
  the random-field heisenberg chain},\ }}\href {\doibase
  10.1103/PhysRevB.91.081103} {\bibfield  {journal} {\bibinfo  {journal} {Phys.
  Rev. B}\ }\textbf {\bibinfo {volume} {91}},\ \bibinfo {pages} {081103}
  (\bibinfo {year} {2015})}\BibitemShut {NoStop}%
\bibitem [{\citenamefont {{Vosk}}\ \emph {et~al.}(2014)\citenamefont {{Vosk}},
  \citenamefont {{Huse}},\ and\ \citenamefont {{Altman}}}]{AltmanRG14}%
  \BibitemOpen
  \bibfield  {author} {\bibinfo {author} {\bibfnamefont {R.}~\bibnamefont
  {{Vosk}}}, \bibinfo {author} {\bibfnamefont {D.~A.}\ \bibnamefont {{Huse}}},
  \ and\ \bibinfo {author} {\bibfnamefont {E.}~\bibnamefont {{Altman}}},\
  }\bibfield  {title} {\emph {\bibinfo {title} {{Theory of the many-body
  localization transition in one dimensional systems}},\ }}\href@noop {}
  {\bibfield  {journal} {\bibinfo  {journal} {ArXiv e-prints}\ } (\bibinfo
  {year} {2014})},\ \Eprint {http://arxiv.org/abs/1412.3117} {arXiv:1412.3117
  [cond-mat.dis-nn]} \BibitemShut {NoStop}%
\bibitem [{\citenamefont {{Potter}}\ \emph {et~al.}(2015)\citenamefont
  {{Potter}}, \citenamefont {{Vasseur}},\ and\ \citenamefont
  {{Parameswaran}}}]{Potter15}%
  \BibitemOpen
  \bibfield  {author} {\bibinfo {author} {\bibfnamefont {A.~C.}\ \bibnamefont
  {{Potter}}}, \bibinfo {author} {\bibfnamefont {R.}~\bibnamefont {{Vasseur}}},
  \ and\ \bibinfo {author} {\bibfnamefont {S.~A.}\ \bibnamefont
  {{Parameswaran}}},\ }\bibfield  {title} {\emph {\bibinfo {title} {{Universal
  properties of many-body delocalization transitions}},\ }}\href@noop {}
  {\bibfield  {journal} {\bibinfo  {journal} {ArXiv e-prints}\ } (\bibinfo
  {year} {2015})},\ \Eprint {http://arxiv.org/abs/1501.03501} {arXiv:1501.03501
  [cond-mat.dis-nn]} \BibitemShut {NoStop}%
\bibitem [{\citenamefont {{Goold}}\ \emph {et~al.}(2015)\citenamefont
  {{Goold}}, \citenamefont {{Clark}}, \citenamefont {{Gogolin}}, \citenamefont
  {{Eisert}}, \citenamefont {{Scardicchio}},\ and\ \citenamefont
  {{Silva}}}]{Goold15}%
  \BibitemOpen
  \bibfield  {author} {\bibinfo {author} {\bibfnamefont {J.}~\bibnamefont
  {{Goold}}}, \bibinfo {author} {\bibfnamefont {S.~R.}\ \bibnamefont
  {{Clark}}}, \bibinfo {author} {\bibfnamefont {C.}~\bibnamefont {{Gogolin}}},
  \bibinfo {author} {\bibfnamefont {J.}~\bibnamefont {{Eisert}}}, \bibinfo
  {author} {\bibfnamefont {A.}~\bibnamefont {{Scardicchio}}}, \ and\ \bibinfo
  {author} {\bibfnamefont {A.}~\bibnamefont {{Silva}}},\ }\bibfield  {title}
  {\emph {\bibinfo {title} {{Total correlations of the diagonal ensemble herald
  the many-body localization transition}},\ }}\href@noop {} {\bibfield
  {journal} {\bibinfo  {journal} {ArXiv e-prints}\ } (\bibinfo {year}
  {2015})},\ \Eprint {http://arxiv.org/abs/1504.06872} {arXiv:1504.06872
  [cond-mat.dis-nn]} \BibitemShut {NoStop}%
\bibitem [{\citenamefont {De~Luca}\ and\ \citenamefont
  {Scardicchio}(2013)}]{Luca13}%
  \BibitemOpen
  \bibfield  {author} {\bibinfo {author} {\bibfnamefont {A.}~\bibnamefont
  {De~Luca}}\ and\ \bibinfo {author} {\bibfnamefont {A.}~\bibnamefont
  {Scardicchio}},\ }\bibfield  {title} {\emph {\bibinfo {title} {Ergodicity
  breaking in a model showing many-body localization},\ }}\href {\doibase
  10.1209/0295-5075/101/37003} {\bibfield  {journal} {\bibinfo  {journal}
  {Europhys. Lett.}\ }\textbf {\bibinfo {volume} {101}},\ \bibinfo {pages}
  {37003} (\bibinfo {year} {2013})}\BibitemShut {NoStop}%
\bibitem [{\citenamefont {{Agarwal}}\ \emph {et~al.}(2014)\citenamefont
  {{Agarwal}}, \citenamefont {{Gopalakrishnan}}, \citenamefont {{Knap}},
  \citenamefont {{Mueller}},\ and\ \citenamefont {{Demler}}}]{Demler14}%
  \BibitemOpen
  \bibfield  {author} {\bibinfo {author} {\bibfnamefont {K.}~\bibnamefont
  {{Agarwal}}}, \bibinfo {author} {\bibfnamefont {S.}~\bibnamefont
  {{Gopalakrishnan}}}, \bibinfo {author} {\bibfnamefont {M.}~\bibnamefont
  {{Knap}}}, \bibinfo {author} {\bibfnamefont {M.}~\bibnamefont {{Mueller}}}, \
  and\ \bibinfo {author} {\bibfnamefont {E.}~\bibnamefont {{Demler}}},\
  }\bibfield  {title} {\emph {\bibinfo {title} {{Anomalous diffusion and
  Griffiths effects near the many-body localization transition}},\ }}\href@noop
  {} {\bibfield  {journal} {\bibinfo  {journal} {ArXiv e-prints}\ } (\bibinfo
  {year} {2014})},\ \Eprint {http://arxiv.org/abs/1408.3413} {arXiv:1408.3413
  [cond-mat.dis-nn]} \BibitemShut {NoStop}%
\bibitem [{\citenamefont {{Torres-Herrera}}\ and\ \citenamefont
  {{Santos}}(2015)}]{Santos15}%
  \BibitemOpen
  \bibfield  {author} {\bibinfo {author} {\bibfnamefont {E.~J.}\ \bibnamefont
  {{Torres-Herrera}}}\ and\ \bibinfo {author} {\bibfnamefont {L.~F.}\
  \bibnamefont {{Santos}}},\ }\bibfield  {title} {\emph {\bibinfo {title}
  {{Dynamics at the Many-Body Localization Transition}},\ }}\href@noop {}
  {\bibfield  {journal} {\bibinfo  {journal} {ArXiv e-prints}\ } (\bibinfo
  {year} {2015})},\ \Eprint {http://arxiv.org/abs/1501.05662} {arXiv:1501.05662
  [cond-mat.dis-nn]} \BibitemShut {NoStop}%
\bibitem [{\citenamefont {{Serbyn}}\ \emph {et~al.}(2015)\citenamefont
  {{Serbyn}}, \citenamefont {{Papic}},\ and\ \citenamefont
  {{Abanin}}}]{Serbyn15}%
  \BibitemOpen
  \bibfield  {author} {\bibinfo {author} {\bibfnamefont {M.}~\bibnamefont
  {{Serbyn}}}, \bibinfo {author} {\bibfnamefont {Z.}~\bibnamefont {{Papic}}}, \
  and\ \bibinfo {author} {\bibfnamefont {D.~A.}\ \bibnamefont {{Abanin}}},\
  }\bibfield  {title} {\emph {\bibinfo {title} {{A criterion for many-body
  localization-delocalization phase transition}},\ }}\href@noop {} {\bibfield
  {journal} {\bibinfo  {journal} {ArXiv e-prints}\ } (\bibinfo {year}
  {2015})},\ \Eprint {http://arxiv.org/abs/1507.01635} {arXiv:1507.01635
  [cond-mat.dis-nn]} \BibitemShut {NoStop}%
\bibitem [{\citenamefont {Izrailev}(1989)}]{Izrailev89}%
  \BibitemOpen
  \bibfield  {author} {\bibinfo {author} {\bibfnamefont {F.~M.}\ \bibnamefont
  {Izrailev}},\ }\bibfield  {title} {\emph {\bibinfo {title} {Intermediate
  statistics of the quasi-energy spectrum and quantum localisation of classical
  chaos},\ }}\href {http://stacks.iop.org/0305-4470/22/i=7/a=017} {\bibfield
  {journal} {\bibinfo  {journal} {Journal of Physics A: Mathematical and
  General}\ }\textbf {\bibinfo {volume} {22}},\ \bibinfo {pages} {865}
  (\bibinfo {year} {1989})}\BibitemShut {NoStop}%
\bibitem [{\citenamefont {Lenz}\ and\ \citenamefont {Haake}(1991)}]{Haake91}%
  \BibitemOpen
  \bibfield  {author} {\bibinfo {author} {\bibfnamefont {G.}~\bibnamefont
  {Lenz}}\ and\ \bibinfo {author} {\bibfnamefont {F.}~\bibnamefont {Haake}},\
  }\bibfield  {title} {\emph {\bibinfo {title} {Reliability of small matrices
  for large spectra with nonuniversal fluctuations},\ }}\href {\doibase
  10.1103/PhysRevLett.67.1} {\bibfield  {journal} {\bibinfo  {journal} {Phys.
  Rev. Lett.}\ }\textbf {\bibinfo {volume} {67}},\ \bibinfo {pages} {1}
  (\bibinfo {year} {1991})}\BibitemShut {NoStop}%
\bibitem [{\citenamefont {Caurier}\ \emph {et~al.}(1990)\citenamefont
  {Caurier}, \citenamefont {Grammaticos},\ and\ \citenamefont
  {Ramani}}]{Caurier}%
  \BibitemOpen
  \bibfield  {author} {\bibinfo {author} {\bibfnamefont {E.}~\bibnamefont
  {Caurier}}, \bibinfo {author} {\bibfnamefont {B.}~\bibnamefont
  {Grammaticos}}, \ and\ \bibinfo {author} {\bibfnamefont {A.}~\bibnamefont
  {Ramani}},\ }\bibfield  {title} {\emph {\bibinfo {title} {Level repulsion
  near integrability: a random matrix analogy},\ }}\href
  {http://stacks.iop.org/0305-4470/23/i=21/a=029} {\bibfield  {journal}
  {\bibinfo  {journal} {Journal of Physics A: Mathematical and General}\
  }\textbf {\bibinfo {volume} {23}},\ \bibinfo {pages} {4903} (\bibinfo {year}
  {1990})}\BibitemShut {NoStop}%
\bibitem [{\citenamefont {Shklovskii}\ \emph {et~al.}(1993)\citenamefont
  {Shklovskii}, \citenamefont {Shapiro}, \citenamefont {Sears}, \citenamefont
  {Lambrianides},\ and\ \citenamefont {Shore}}]{Shklovskii93}%
  \BibitemOpen
  \bibfield  {author} {\bibinfo {author} {\bibfnamefont {B.~I.}\ \bibnamefont
  {Shklovskii}}, \bibinfo {author} {\bibfnamefont {B.}~\bibnamefont {Shapiro}},
  \bibinfo {author} {\bibfnamefont {B.~R.}\ \bibnamefont {Sears}}, \bibinfo
  {author} {\bibfnamefont {P.}~\bibnamefont {Lambrianides}}, \ and\ \bibinfo
  {author} {\bibfnamefont {H.~B.}\ \bibnamefont {Shore}},\ }\bibfield  {title}
  {\emph {\bibinfo {title} {Statistics of spectra of disordered systems near
  the metal-insulator transition},\ }}\href {\doibase
  10.1103/PhysRevB.47.11487} {\bibfield  {journal} {\bibinfo  {journal} {Phys.
  Rev. B}\ }\textbf {\bibinfo {volume} {47}},\ \bibinfo {pages} {11487}
  (\bibinfo {year} {1993})}\BibitemShut {NoStop}%
\bibitem [{\citenamefont {Evers}\ and\ \citenamefont
  {Mirlin}(2008)}]{MirlinRMP}%
  \BibitemOpen
  \bibfield  {author} {\bibinfo {author} {\bibfnamefont {F.}~\bibnamefont
  {Evers}}\ and\ \bibinfo {author} {\bibfnamefont {A.~D.}\ \bibnamefont
  {Mirlin}},\ }\bibfield  {title} {\emph {\bibinfo {title} {Anderson
  transitions},\ }}\href {\doibase 10.1103/RevModPhys.80.1355} {\bibfield
  {journal} {\bibinfo  {journal} {Rev. Mod. Phys.}\ }\textbf {\bibinfo {volume}
  {80}},\ \bibinfo {pages} {1355} (\bibinfo {year} {2008})}\BibitemShut
  {NoStop}%
\bibitem [{\citenamefont {Mirlin}(2000)}]{MirlinRPP}%
  \BibitemOpen
  \bibfield  {author} {\bibinfo {author} {\bibfnamefont {A.~D.}\ \bibnamefont
  {Mirlin}},\ }\bibfield  {title} {\emph {\bibinfo {title} {Statistics of
  energy levels and eigenfunctions in disordered systems},\ }}\href {\doibase
  http://dx.doi.org/10.1016/S0370-1573(99)00091-5} {\bibfield  {journal}
  {\bibinfo  {journal} {Physics Reports}\ }\textbf {\bibinfo {volume} {326}},\
  \bibinfo {pages} {259 } (\bibinfo {year} {2000})}\BibitemShut {NoStop}%
\bibitem [{\citenamefont {Modak}\ \emph {et~al.}(2014)\citenamefont {Modak},
  \citenamefont {Mukerjee},\ and\ \citenamefont {Ramaswamy}}]{Modak}%
  \BibitemOpen
  \bibfield  {author} {\bibinfo {author} {\bibfnamefont {R.}~\bibnamefont
  {Modak}}, \bibinfo {author} {\bibfnamefont {S.}~\bibnamefont {Mukerjee}}, \
  and\ \bibinfo {author} {\bibfnamefont {S.}~\bibnamefont {Ramaswamy}},\
  }\bibfield  {title} {\emph {\bibinfo {title} {Universal power law in
  crossover from integrability to quantum chaos},\ }}\href {\doibase
  10.1103/PhysRevB.90.075152} {\bibfield  {journal} {\bibinfo  {journal} {Phys.
  Rev. B}\ }\textbf {\bibinfo {volume} {90}},\ \bibinfo {pages} {075152}
  (\bibinfo {year} {2014})}\BibitemShut {NoStop}%
\bibitem [{\citenamefont {Oganesyan}\ and\ \citenamefont
  {Huse}(2007)}]{OganesyanHuse}%
  \BibitemOpen
  \bibfield  {author} {\bibinfo {author} {\bibfnamefont {V.}~\bibnamefont
  {Oganesyan}}\ and\ \bibinfo {author} {\bibfnamefont {D.~A.}\ \bibnamefont
  {Huse}},\ }\bibfield  {title} {\emph {\bibinfo {title} {Localization of
  interacting fermions at high temperature},\ }}\href {\doibase
  10.1103/PhysRevB.75.155111} {\bibfield  {journal} {\bibinfo  {journal} {Phys.
  Rev. B}\ }\textbf {\bibinfo {volume} {75}},\ \bibinfo {pages} {155111}
  (\bibinfo {year} {2007})}\BibitemShut {NoStop}%
\bibitem [{\citenamefont {Avishai}\ \emph {et~al.}(2002)\citenamefont
  {Avishai}, \citenamefont {Richert},\ and\ \citenamefont
  {Berkovits}}]{Berkovits02}%
  \BibitemOpen
  \bibfield  {author} {\bibinfo {author} {\bibfnamefont {Y.}~\bibnamefont
  {Avishai}}, \bibinfo {author} {\bibfnamefont {J.}~\bibnamefont {Richert}}, \
  and\ \bibinfo {author} {\bibfnamefont {R.}~\bibnamefont {Berkovits}},\
  }\bibfield  {title} {\emph {\bibinfo {title} {Level statistics in a
  heisenberg chain with random magnetic field},\ }}\href {\doibase
  10.1103/PhysRevB.66.052416} {\bibfield  {journal} {\bibinfo  {journal} {Phys.
  Rev. B}\ }\textbf {\bibinfo {volume} {66}},\ \bibinfo {pages} {052416}
  (\bibinfo {year} {2002})}\BibitemShut {NoStop}%
\bibitem [{\citenamefont {Modak}\ and\ \citenamefont
  {Mukerjee}(2014)}]{Modak14}%
  \BibitemOpen
  \bibfield  {author} {\bibinfo {author} {\bibfnamefont {R.}~\bibnamefont
  {Modak}}\ and\ \bibinfo {author} {\bibfnamefont {S.}~\bibnamefont
  {Mukerjee}},\ }\bibfield  {title} {\emph {\bibinfo {title} {Finite size
  scaling in crossover among different random matrix ensembles in microscopic
  lattice models},\ }}\href {http://stacks.iop.org/1367-2630/16/i=9/a=093016}
  {\bibfield  {journal} {\bibinfo  {journal} {New Journal of Physics}\ }\textbf
  {\bibinfo {volume} {16}},\ \bibinfo {pages} {093016} (\bibinfo {year}
  {2014})}\BibitemShut {NoStop}%
\bibitem [{\citenamefont {Dyson}(1962)}]{Dyson62}%
  \BibitemOpen
  \bibfield  {author} {\bibinfo {author} {\bibfnamefont {F.~J.}\ \bibnamefont
  {Dyson}},\ }\bibfield  {title} {\emph {\bibinfo {title} {{A Brownian-Motion
  Model for the Eigenvalues of a Random Matrix}},\ }}\href {\doibase
  http://dx.doi.org/10.1063/1.1703862} {\bibfield  {journal} {\bibinfo
  {journal} {Journal of Mathematical Physics}\ }\textbf {\bibinfo {volume}
  {3}},\ \bibinfo {pages} {1191} (\bibinfo {year} {1962})}\BibitemShut
  {NoStop}%
\bibitem [{\citenamefont {Chalker}\ \emph {et~al.}(1996)\citenamefont
  {Chalker}, \citenamefont {Lerner},\ and\ \citenamefont {Smith}}]{ChalkerRW}%
  \BibitemOpen
  \bibfield  {author} {\bibinfo {author} {\bibfnamefont {J.~T.}\ \bibnamefont
  {Chalker}}, \bibinfo {author} {\bibfnamefont {I.~V.}\ \bibnamefont {Lerner}},
  \ and\ \bibinfo {author} {\bibfnamefont {R.~A.}\ \bibnamefont {Smith}},\
  }\bibfield  {title} {\emph {\bibinfo {title} {Random walks through the
  ensemble: Linking spectral statistics with wave-function correlations in
  disordered metals},\ }}\href {\doibase 10.1103/PhysRevLett.77.554} {\bibfield
   {journal} {\bibinfo  {journal} {Phys. Rev. Lett.}\ }\textbf {\bibinfo
  {volume} {77}},\ \bibinfo {pages} {554} (\bibinfo {year} {1996})}\BibitemShut
  {NoStop}%
\bibitem [{\citenamefont {Nandkishore}\ \emph {et~al.}(2014)\citenamefont
  {Nandkishore}, \citenamefont {Gopalakrishnan},\ and\ \citenamefont
  {Huse}}]{Rahul14}%
  \BibitemOpen
  \bibfield  {author} {\bibinfo {author} {\bibfnamefont {R.}~\bibnamefont
  {Nandkishore}}, \bibinfo {author} {\bibfnamefont {S.}~\bibnamefont
  {Gopalakrishnan}}, \ and\ \bibinfo {author} {\bibfnamefont {D.~A.}\
  \bibnamefont {Huse}},\ }\bibfield  {title} {\emph {\bibinfo {title} {Spectral
  features of a many-body-localized system weakly coupled to a bath},\ }}\href
  {\doibase 10.1103/PhysRevB.90.064203} {\bibfield  {journal} {\bibinfo
  {journal} {Phys. Rev. B}\ }\textbf {\bibinfo {volume} {90}},\ \bibinfo
  {pages} {064203} (\bibinfo {year} {2014})}\BibitemShut {NoStop}%
\bibitem [{\citenamefont {Bogomolny}\ \emph {et~al.}(1999)\citenamefont
  {Bogomolny}, \citenamefont {Gerland},\ and\ \citenamefont
  {Schmit}}]{Bogomolny99}%
  \BibitemOpen
  \bibfield  {author} {\bibinfo {author} {\bibfnamefont {E.~B.}\ \bibnamefont
  {Bogomolny}}, \bibinfo {author} {\bibfnamefont {U.}~\bibnamefont {Gerland}},
  \ and\ \bibinfo {author} {\bibfnamefont {C.}~\bibnamefont {Schmit}},\
  }\bibfield  {title} {\emph {\bibinfo {title} {Models of intermediate spectral
  statistics},\ }}\href {\doibase 10.1103/PhysRevE.59.R1315} {\bibfield
  {journal} {\bibinfo  {journal} {Phys. Rev. E}\ }\textbf {\bibinfo {volume}
  {59}},\ \bibinfo {pages} {R1315} (\bibinfo {year} {1999})}\BibitemShut
  {NoStop}%
\bibitem [{\citenamefont {Chalker}\ and\ \citenamefont
  {Daniell}(1988)}]{ChalkerDaniell}%
  \BibitemOpen
  \bibfield  {author} {\bibinfo {author} {\bibfnamefont {J.~T.}\ \bibnamefont
  {Chalker}}\ and\ \bibinfo {author} {\bibfnamefont {G.~J.}\ \bibnamefont
  {Daniell}},\ }\bibfield  {title} {\emph {\bibinfo {title} {Scaling,
  diffusion, and the integer quantized hall effect},\ }}\href {\doibase
  10.1103/PhysRevLett.61.593} {\bibfield  {journal} {\bibinfo  {journal} {Phys.
  Rev. Lett.}\ }\textbf {\bibinfo {volume} {61}},\ \bibinfo {pages} {593}
  (\bibinfo {year} {1988})}\BibitemShut {NoStop}%
\bibitem [{sup()}]{suppmat}%
  \BibitemOpen
  \href@noop {} {}\bibinfo {note} {See Supplemental Material}\BibitemShut
  {NoStop}%
\bibitem [{\citenamefont {Cuevas}\ and\ \citenamefont
  {Kravtsov}(2007)}]{KravtsovEigenfunctionCorrelation}%
  \BibitemOpen
  \bibfield  {author} {\bibinfo {author} {\bibfnamefont {E.}~\bibnamefont
  {Cuevas}}\ and\ \bibinfo {author} {\bibfnamefont {V.~E.}\ \bibnamefont
  {Kravtsov}},\ }\bibfield  {title} {\emph {\bibinfo {title} {Two-eigenfunction
  correlation in a multifractal metal and insulator},\ }}\href {\doibase
  10.1103/PhysRevB.76.235119} {\bibfield  {journal} {\bibinfo  {journal} {Phys.
  Rev. B}\ }\textbf {\bibinfo {volume} {76}},\ \bibinfo {pages} {235119}
  (\bibinfo {year} {2007})}\BibitemShut {NoStop}%
\bibitem [{\citenamefont {Kravtsov}\ and\ \citenamefont
  {Lerner}(1995)}]{KravtsovPlasma}%
  \BibitemOpen
  \bibfield  {author} {\bibinfo {author} {\bibfnamefont {V.~E.}\ \bibnamefont
  {Kravtsov}}\ and\ \bibinfo {author} {\bibfnamefont {I.~V.}\ \bibnamefont
  {Lerner}},\ }\bibfield  {title} {\emph {\bibinfo {title} {Effective plasma
  model for the level correlations at the mobility edge},\ }}\href
  {http://stacks.iop.org/0305-4470/28/i=13/a=008} {\bibfield  {journal}
  {\bibinfo  {journal} {Journal of Physics A: Mathematical and General}\
  }\textbf {\bibinfo {volume} {28}},\ \bibinfo {pages} {3623} (\bibinfo {year}
  {1995})}\BibitemShut {NoStop}%
\bibitem [{\citenamefont {Kravtsov}\ and\ \citenamefont
  {Muttalib}(1997)}]{KravtsovNewClass}%
  \BibitemOpen
  \bibfield  {author} {\bibinfo {author} {\bibfnamefont {V.~E.}\ \bibnamefont
  {Kravtsov}}\ and\ \bibinfo {author} {\bibfnamefont {K.~A.}\ \bibnamefont
  {Muttalib}},\ }\bibfield  {title} {\emph {\bibinfo {title} {New class of
  random matrix ensembles with multifractal eigenvectors},\ }}\href {\doibase
  10.1103/PhysRevLett.79.1913} {\bibfield  {journal} {\bibinfo  {journal}
  {Phys. Rev. Lett.}\ }\textbf {\bibinfo {volume} {79}},\ \bibinfo {pages}
  {1913} (\bibinfo {year} {1997})}\BibitemShut {NoStop}%
\bibitem [{\citenamefont {Muttalib}\ \emph {et~al.}(1993)\citenamefont
  {Muttalib}, \citenamefont {Chen}, \citenamefont {Ismail},\ and\ \citenamefont
  {Nicopoulos}}]{MuttalibNewFamily}%
  \BibitemOpen
  \bibfield  {author} {\bibinfo {author} {\bibfnamefont {K.~A.}\ \bibnamefont
  {Muttalib}}, \bibinfo {author} {\bibfnamefont {Y.}~\bibnamefont {Chen}},
  \bibinfo {author} {\bibfnamefont {M.~E.~H.}\ \bibnamefont {Ismail}}, \ and\
  \bibinfo {author} {\bibfnamefont {V.~N.}\ \bibnamefont {Nicopoulos}},\
  }\bibfield  {title} {\emph {\bibinfo {title} {New family of unitary random
  matrices},\ }}\href {\doibase 10.1103/PhysRevLett.71.471} {\bibfield
  {journal} {\bibinfo  {journal} {Phys. Rev. Lett.}\ }\textbf {\bibinfo
  {volume} {71}},\ \bibinfo {pages} {471} (\bibinfo {year} {1993})}\BibitemShut
  {NoStop}%
\bibitem [{\citenamefont {Garc\'{i}a-Garc\'{i}a}\ and\ \citenamefont
  {Verbaarschot}(2003)}]{VerbaarchotCritical}%
  \BibitemOpen
  \bibfield  {author} {\bibinfo {author} {\bibfnamefont {A.~M.}\ \bibnamefont
  {Garc\'{i}a-Garc\'{i}a}}\ and\ \bibinfo {author} {\bibfnamefont {J.~J.~M.}\
  \bibnamefont {Verbaarschot}},\ }\bibfield  {title} {\emph {\bibinfo {title}
  {Critical statistics in quantum chaos and calogero-sutherland model at finite
  temperature},\ }}\href {\doibase 10.1103/PhysRevE.67.046104} {\bibfield
  {journal} {\bibinfo  {journal} {Phys. Rev. E}\ }\textbf {\bibinfo {volume}
  {67}},\ \bibinfo {pages} {046104} (\bibinfo {year} {2003})}\BibitemShut
  {NoStop}%
\bibitem [{\citenamefont {G\'omez}\ \emph {et~al.}(2002)\citenamefont
  {G\'omez}, \citenamefont {Molina}, \citenamefont {Rela\~no},\ and\
  \citenamefont {Retamosa}}]{Misleading}%
  \BibitemOpen
  \bibfield  {author} {\bibinfo {author} {\bibfnamefont {J.~M.~G.}\
  \bibnamefont {G\'omez}}, \bibinfo {author} {\bibfnamefont {R.~A.}\
  \bibnamefont {Molina}}, \bibinfo {author} {\bibfnamefont {A.}~\bibnamefont
  {Rela\~no}}, \ and\ \bibinfo {author} {\bibfnamefont {J.}~\bibnamefont
  {Retamosa}},\ }\bibfield  {title} {\emph {\bibinfo {title} {Misleading
  signatures of quantum chaos},\ }}\href {\doibase 10.1103/PhysRevE.66.036209}
  {\bibfield  {journal} {\bibinfo  {journal} {Phys. Rev. E}\ }\textbf {\bibinfo
  {volume} {66}},\ \bibinfo {pages} {036209} (\bibinfo {year}
  {2002})}\BibitemShut {NoStop}%
\end{thebibliography}
\end{document}